\documentclass[printer]{gPHT2e}
\usepackage{rotating}
\usepackage{cite}
\begin{document}
\setlength{\oddsidemargin}{4.0cm}
\setlength{\evensidemargin}{4.0cm}
\setlength{\topmargin}{4.0cm}
\doi{10.1080/014115940xxxxxxxxxxxx}
 \issn{1029-0338}
 \issnp{0141-1594} \jvol{?} \jnum{?} \jyear{200?}
 \jmonth{?}

\markboth{%
Magnetic interactions in Cu-containing heterospin polymer}{%
A. V. Postnikov et al.}

\title{Magnetic interactions in Cu-containing heterospin polymer}

\author{A. V. POSTNIKOV$\dagger\ddagger$$^{\ast}$
\thanks{$^{\ast}$Corresponding author. E-mail: apostnik@uos.de},  
A. V. GALAKHOV$\ddagger$ and 
S. BL\"UGEL$\dagger$ \\
$\dagger$
Institut f\"ur Festk\"orperforschung, Forschungszentrum J\"ulich,
D-52425 J\"ulich, Germany \\
$\ddagger$
Institute of Metal Physics, S. Kowalewskoj 18, Yekaterinburg 620219, Russia
}   
\received{received in June 2005}
\maketitle

\begin{abstract}
The electronic structure of Cu hexafluoroacetylacetonate, crystallized
with a stable nitronyl nitroxide radical [Ovcharenko \emph{et al.},
Russ. Chem. Bull. {\bf 53}, 2406 (2004)], is calculated from first
principles within the density functional theory using the SIESTA method,
in two magnetic configurations reflecting parallel or antiparallel
setting of $S=1/2$ spins of Cu(II) ions to those of organic radicals.
For a given (high-temperature) crystal structure, the interaction is
found to be predominanty antiferromagnetic, and its magnitude estimated
to be 67 cm$^{-1}$. This preference is discussed in terms of
calculated electronic properties (densities of states, molecular orbitals).
\end{abstract}

\section{Introduction}
Over last decades, a large variety of metalloorganic materials 
has been synthesized, many of them having non-trivial magnetic properties.
One often denotes them generally as ``molecular magnets'', notwithstanding 
their affiliation to one or another of two large groups
(see, e.g., \cite{Kahn-book,Mol_Magnets}).
The first is referred to as ``single molecule magnets'', which are
molecular crystals composed of well-defined blocks, stable in a solution
and remaining well spatially separated (by solvent molecules)
upon crystallization. 
Such systems gained much attention due to spectacular manifestations
of essentially quantum mechanical properties (quantum tunneling of
magnetization, spin crossiver) in the behaviour of ensembles of identical
molecules. Another group of systems, although also being often
refered to as molecular magnets, are in fact polymers, because their
molecular building blocks gets connected into effectively infinite
linear chains, or more sophisticated structures. The propeties of
a single molecule cannot be easily specified in such systems, but
a number of other characteristics become relevant, e.g. conductivity
and long-range magnetic order. Among such polymers one finds metallorganic
ferromagnets with respectable values ($>$300~K) 
of Curie temperature \cite{PhilTransA357-1762}.

Further on, the magnetism is not necessarily due to transition metal (TM) ions,
but it can be brought about by a free radical as well. The study of
polymers based on stable free radicals has a long history.
Mixing up TM ions and free radivals in a single polymer system
opens possibilities of sophisticated tuning of magnetic properties.
The chemistry offers certain flexibility in bringing together
the localized spins of TM ions and distributed spins of free radicals
in a controllable way, in a pattern reproducible throughout the crystal.
A recent review to this subject has been offered by Blundell and
Pratt \cite{JPCM16-R771}.

This poses a question to which extent the electronic and magnetic
properties of such heterospin systems can be grasped by a first-principles
calculation. While in general \emph{ab initio} calculation schemes
based on the density functional theory (DFT) work well (up to certain
limitations) in describing the ground-state magnetic structure and 
certain related properties (magetic anisotropy etc., see 
Ref.~\cite{Psik-highlight,MolMag-book} for reviews), heterospin polymer 
systems have certain obstacles
for a calculation. First, many DFT methods found useful
in calculations of single molecular magnets, can work on isolated
molecular fragments only, and do not support periodic boundary conditions.
The same applies to more accurate quantum chemistry calculation schemes.
Second, large number of atoms and lack of symmetry makes 
an \emph{ab initio} calculation with any scheme quite demanding.
However, a successful calculation may give insight into chemical bonding
and its relation with magnetic properties in these heterospin 
compounds. Moreover one can apply certain constraints in the calculation
and force the system into one or another magnetic configuration,
which would help to estimate the typical magnetic interaction energies.

\section{The system of the present study}

We calculate and discuss the electronic structure of
one heterospin substance, based on copper (II) hexafluoroacetylacetonate
Cu(hfac)$_2$, hfac = CF$_3$--C(O)--CH--C(O)--CF$_3$, 
in combination with a stable nitronyl nitroxide radical
4,4,5,5,-tetramethyl-2(1-ethyl-1$H$-pyrazol--4-yl)-imidazoline-3-oxide-1-oxyl,
described e.g. in Ref.~\cite{ZhStCh43-163}. Some properties
of the resulting crystal were addressed in Ref.~\cite{PhSS45-1465}.
During last years a large variety of related heterospin compounds 
has been synthesized\cite{MolPhys100-1107,InCh43-969,ZhStCh43-163},
in which Cu(hfac)$_2$ blocks are connected by spin-labeled pirazoles
(L$^{\rm Me}$, L$^{\rm Et}$, L$^{\rm i-Pr}$, L$^{\rm Bu}$ 
for the above formula with different 1-alkyl substitutions).

\begin{figure}
\centerline{\includegraphics[angle=0,width=13cm]{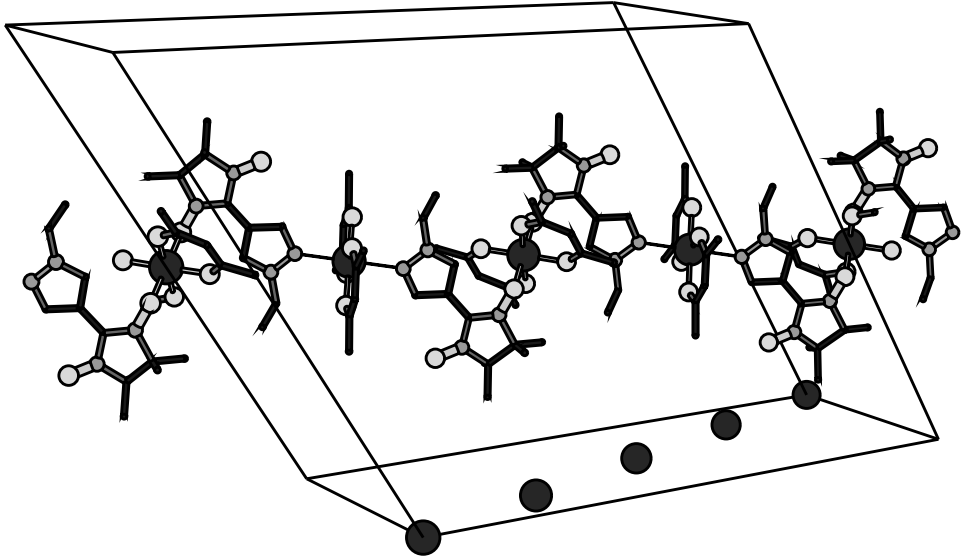}}
\caption{Monoclinic unit cell of Cu(hfac)$_2$$L$. One chain of Cu atoms
with their embedding (hfac)$_2$ fragments and connecting 
radical molecules runs along
the diagonal of the mid-height cutting plane; of an (equivalent) chain
in the basal plane, only Cu atoms are shown. H and F atoms are omitted.
}
\label{fig:cell}
\end{figure}

\begin{figure}
\centerline{\includegraphics[angle=0,width=11cm]{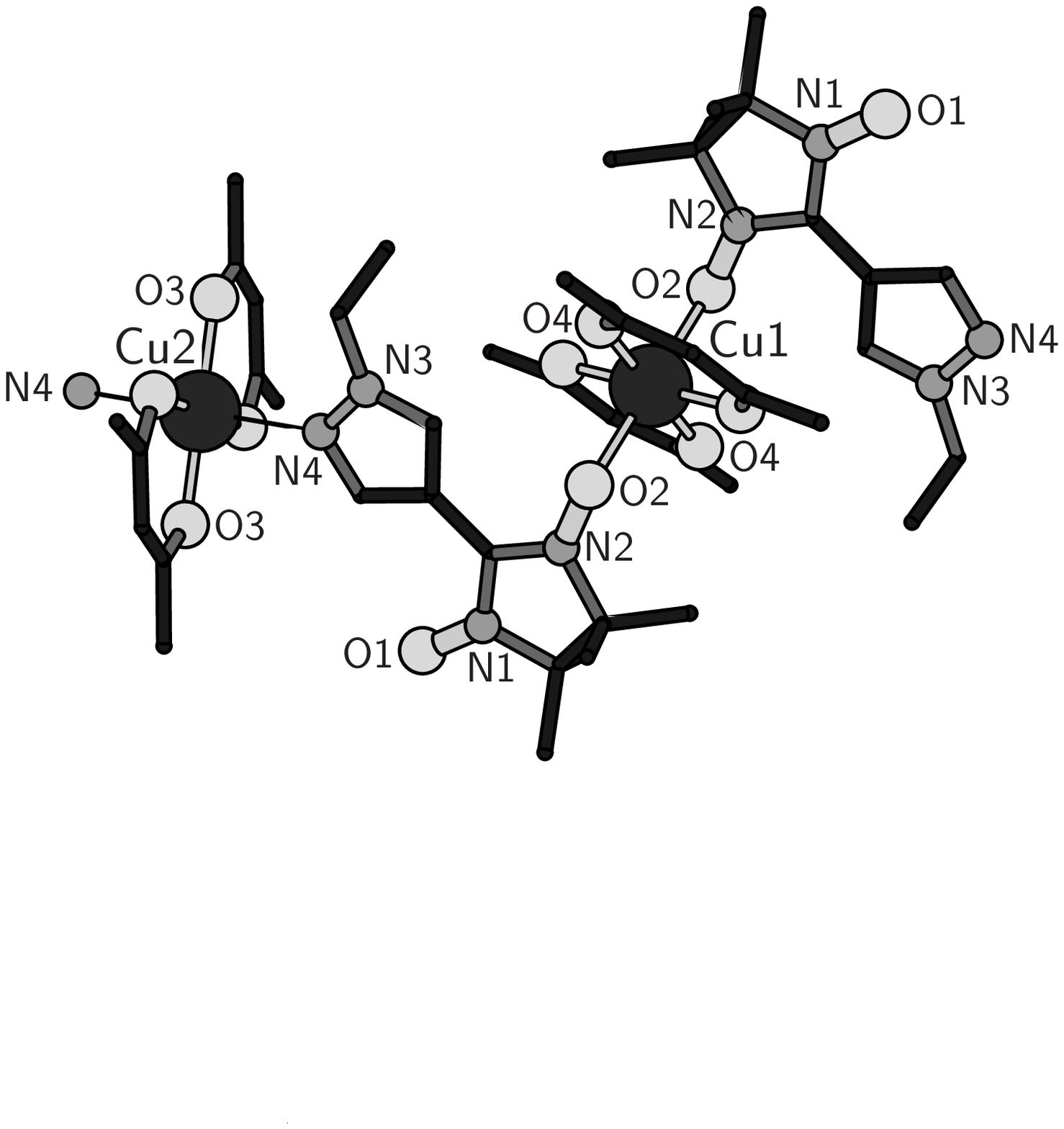}}
\caption{An enlarged view of Cu(hfac)$_2$--L--Cu(hfac)$_2$--L
sequence, making a period of the chain in Fig.~\ref{fig:cell},
with labelling of atoms.
}
\label{fig:unit}
\end{figure}

The compound in question with the nominal formula
C$_{22}$H$_{21}$CuF$_{12}$N$_4$O$_6$, refered to as Cu(hfac)$_2$L$^{\rm Et}$, 
crystallizes in base-centered monoclinic lattice with the space group 
$C2/c$ (Nr. 15) and lattice parameters $a$=30.516 {\AA}, $b$=9.540 {\AA}, 
$c$=25.194 {\AA}, $\beta$=123.96$^{\circ}$. The primitive cell contains 
four above given formula units (264 atoms). The monoclinic unit cell 
(hosting twice as many atoms) is shown in Fig.~\ref{fig:cell}; 
the arrangement and labelling of atoms in the fragments relevant 
for our discussion, -- `head-to-head' coupling of ligands to magnetic 
Cu atoms, embedded in the (hfac)$_2$ blocks, -- in Fig.~\ref{fig:unit}.

\section{Calculation method and setup}

The calculation have been done within the DFT, by the
{\sc Siesta} method and computer code\cite{JPCM14-2745,PRB53-10441,siesta},
using atom-centered and strictly confined numerical 
functions\cite{JPCM8-3859,PRB64-235111} as a basis set
for solving the Kohn--Sham equations. Norm-conserving pseudopotentials
have been constructed according the Troullier--Martins scheme\cite{PRB43-1993},
for the following electron configurations (pseudoization radii in Bohr
are given in brackets): 
C$2s^2$(1.25)$2p^2$(1.25),
N$2s^2$(1.25)$2p^3$(1.25),
O$2s^2$(1.15)$2p^4$(1.15),
F$2s^2$(1.20)$2p^5$(1.20),
Cu$4s^{1.50}$(2.08)$4p^0$(2.08)$3d^{9.50}$(1.79).
The basis set was double-$\zeta$ with polarization orbitals, according
to the notation common in tight-binding schemes (see, e.g.,
Refs.~\cite{JPCM8-3859,PRB64-235111}) for  Cu, N and O, 
and double-$\zeta$ for C, F and H. The energy shift parameter
responsible for the localization of basis functions in {\sc Siesta}
with its `default' value of 20 mRy resulted in the following maximal 
extension of basis functions: 
2.492 {\AA} (H), 2.577 {\AA} (C), 2.265 {\AA} (N),
2.083 {\AA} (O), 1.899 {\AA} (F), 3.308 {\AA} (Cu).
The sampling over the Brillouin zone has been done with 
3$\times$5$\times$2 divisions along the reciprocal cell vectors.
The real-space grid used for the fast Fourier transform in the process
of solving the Poisson equation was generated with the energy cutoff
parameter of beyond 300 Ry, that resulted in 144$\times$96$\times$240
divisions along the lattice vectors.
The exchange-correlation was treated in the generalized gradient approximation 
after Perdew--Burke--Ernzerhof \cite{PRL77-3865}. 

\section{Results and discussion}

We used the coordinates of atoms as determined experimentally
\cite{ZhStCh43-163,Fokin} at room temperature,
without attempting to adjust the structure in the calculation. 
In fact the crystal structure, notably Cu1--O2 and Cu2--N4 bonds,
slightly vary with the temperature\cite{ZhStCh43-163}. Apparently
this gives rise to an abrupt change of the character of magnetic
interaction between Cu1 and radical spin, from predominantly antiferromagnetic
(above $\sim$220 K) to predominantly ferromagnetic (below this temperature),
as the Cu1--O2 distance grows from 2.237~{\AA} (293 K) to 2.260 (188 K) 
to 2.281 (115 K). Ovcharenko \emph{et al.}\cite{ZhStCh43-163} emphasize
that antiferromagnetic coupling in the axially coordinated
\mbox{$>$N$-^{\bm \cdot}$O$-$Cu(II)$-$O$^{\bm \cdot}-$N$<$} groups 
is rather unusual
and comes about due to anomalously short Cu--O distance at higher
temperatures. It was a major objective of our present study to verify 
a tendency for antiferromagnetic cooupling from a first principles
calculation.

\begin{figure}
\centerline{\includegraphics[angle=0,width=12.5cm]{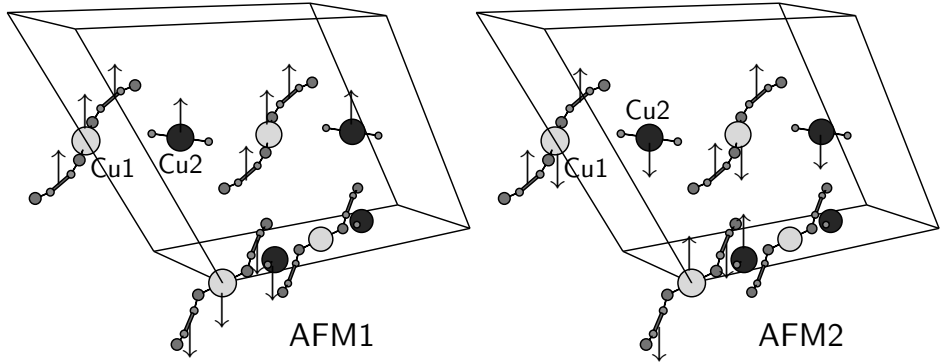}}
\caption{Orientation of magnetic moments in two tested AFM configurations.
Only Cu1 atoms flanked by O--N$\cdots$N--O groups of the radical
and Cu2 with their two N4 neighbours are shown.
}
\label{fig:AFM}
\end{figure}

The calculation can be initialized with different spin configurations
on selected atoms; this leads to a set of solutions,
representing different metastable configurations. Two observations
helped to reduce the amount of work: 
$i)$ the solutions with opposite setting of magnetic moments
of two radicals neighbouring a certain Cu1 atom were impossible
to bring to convergence, thus indicating that a symmetry or antisymmetry
of hybridized wavefunctions with respect to a central Cu1 are strongly
favoured;
$ii)$ the energy difference between solutions which correspond to
various magnetic order between two chains (shown in Fig.~\ref{fig:cell}),
i.e. due to inversion of all spins in one chain, are tiny -- at least
they are not reliably resolvable within the present calculation setup.
This underlines an expected fact that magnetic interaction between two
(spatially well separated) chains are negligibly small. Therefore
we imposed antiparallel setting of spin moments within one chain
with respect to the other, and fixed
the total spin moment of the unit cell to zero, which helped to improve
the stability and convergency of calculation.
This leaves us with several options of mutually setting spins of Cu1, 
two radicals coupled to it, and Cu2. We keep two most stable solutions,
shown in Fig.~\ref{fig:AFM}, and discuss their electronic structure
in more detail.

\begin{table}
\tbl{Partial charges (Mulliken populations) $Q$ ($e$) 
and local magnetic moments $M$ ($\mu_{\rm B}$)
in two most stable magnetic configurations. Four oxygen atoms in each
Cu(hfac)$_2$ fragment, labeled O3 and O4 in Fig.~\ref{fig:unit},
are in reality subdivided into two inequivalent sites,
according to inversion symmetry relative to Cu centers.}
 {\begin{tabular}{@{}cr@{.}lr@{.}lcr@{.}lr@{.}lcr@{.}lr@{.}lr@{.}lr@{.}l}
\toprule
 & \multicolumn{2}{c}{Cu1} & \multicolumn{2}{c}{O4} &&
   \multicolumn{2}{c}{Cu2} & \multicolumn{2}{c}{O3} &&
   \multicolumn{2}{c}{O1}  & \multicolumn{2}{c}{N1} & 
   \multicolumn{2}{c}{N2}  & \multicolumn{2}{c}{O2} \\
\colrule
 \multicolumn{6}{c}{} & \multicolumn{4}{c}{AFM1} \\
 $Q$: & 10&64 & \multicolumn{2}{c}{6.15} &&
        10&67 & \multicolumn{2}{c}{6.14--6.16} &&
         6&10 &  4&93 &  4&90 & 6&15 \\
 $M$: &  0&51 & \multicolumn{2}{c}{0.07} &&
         0&57 & \multicolumn{2}{c}{0.08--0.09} &&
         0&29 &  0&21 &  0&20 & 0&19 \\
\colrule
 \multicolumn{6}{c}{} & \multicolumn{4}{c}{AFM2} \\
 $Q$: & 10&65 & \multicolumn{2}{c}{6.15--6.16} && 
        10&67 & \multicolumn{2}{c}{6.15--6.16} &&
         6&10 &  4&93 &  4&90 & 6&15 \\
 $M$: &  0&45 & \multicolumn{2}{c}{0.08--0.04} && 
         0&57 & \multicolumn{2}{c}{0.07--0.08} &&
         0&30 &  0&21 &  0&21 & 0&18 \\
\botrule
\end{tabular}}
\label{tab:charge}
\end{table}

The AFM2 case involves the inversion of magnetic moments of all Cu atoms, 
as compared to AFM1. This leaves the local electronic structure
only slightly affected; the Mulliken charges and local magnetic moments
(Table~\ref{tab:charge}) are the same, within 0.01, for AFM1 and AFM2. 
One finds 1~$\mu_{\rm B}$ distributed over the radical (referred to as R
in the following); similarly, the spin $S$=1/2 can be associated with 
each Cu site, even as in reality the magnetic density spills over its 
neighbouring atoms of the (hfac)$_2$ fragment. 
We found a band dispersion over sampled points in the Brilluoin zone
to be within 0.1 meV, preserving a band gap of $\sim$16 mRy in AFM1
and $\sim$36 meV in the AFM2 configuration. Although not a strict criterion, 
a larger band gap is often an indication of a more stable configuration 
among competitive ones. Indeed, the total energy is lower in the AFM2 case 
by 35 meV, per unit cell of 256 atoms, which contains two Cu2--L--Cu1--L units.
It could be expected that
the experimental band gap be larger, because the molecular orbitals
just below and above the gap are those with large Cu contribution,
and it is known that an inclusion of correlation effects within the 
(quite localized, in this case) Cu$3d$ shell
beyond the conventional GGA would tend to move these states apart.

Assuming the Cu1--R magnetic interaction
as dominant and those involving Cu2 as comparably small, one can map
the above energy difference onto the Heisenberg model. 
There are several ways to extract magnetic interaction parameters
from a first-principles calculation, e.g., in a form consistent with 
spin-fluctuation theories, in terms of Green's function elements
\cite{JPF14-L125,JMMM67-65,PhysB237-336},
or via non-local susceptibility \cite{PRB52-R5467}.
The difficulty in the present case would be a delocalized character
of a spin over four atoms (O1--N1$\cdots$N2--O2) of the radical, and a certain
ambiguity in defining the corresponding matrix elements of the Green's
function and/or susceptibility. A less conceptually problematic,
although quite `old-fashioned', is comparison of total energies in
several representative magnetic configurations. A prerequisite for
this approach to be meaningful is that the involved spins retain their
`identity' (i.e., magnitude, localization etc.) in the configurations
to be compared. This is indeed the case in the compound under study,
as follows from the inspection of local densities of states (DOS,
shown below) and spatial spin densities.

Given ambiguities in the definitions of sign and prefacor of the
Heisenberg model, we introduce the magnetic interaction parameter
consistently with Ref.~\cite{InCh43-969}, notably Eq.~(2) therein
for the special case of interaction within the 
\mbox{$>$N$-^{\bm \cdot}$O$-$Cu(II)$-$O$^{\bm \cdot}-$N$<$}
`exchange cluster': 
\[
{\cal H} = -2 J {\bm S}_{\rm Cu}{\bm S}' +
        \mbox{(Zeeman term)} + \mbox{(interaction with more distant ions)}\,,
\]
where ${\bm S}'$ is the total spin of \emph{two} radicals flanking Cu1.
Mapping this onto total energy values of our calculation yields:
\[
E_{\rm AFM1}-E_{\rm AFM2} = -16 J S_{\rm Cu} S_{\rm R}\,,
\]
whence (with $S_{\rm R}$=1/2 for the spin of one radical) 
$-\!J$ = 8.3~meV = 96~K = 67~cm$^{-1}$. 
This rough estimate of (antiferromagnetic)
coupling between a Cu ion and the radical can be looked at as an
upper-bonded value: an inclusion of intraatomic correlation effect
beyond the conventional GGA may scale down the energy difference
between magnetic configurations by a factor as large as 3--4, depending
on a system, according to
a previous experience for other molecular magnets\cite{PRB65-184435,Novo04-Fe}.
Moreover, an explicit inclusion of interactions with Cu2 into the model
would slightly renormalize the Cu1--R interaction further down.

We could not find an experimental value to compare with, e.g. that
extracted from a Heisenberg-model fit to magnetic measurements data; 
the value $J$=26~cm$^{-1}$ given in Table 3 of Ref.~\cite{ZhStCh43-163} 
stems from the fit over low-temperature range, where 
exchange interaction within the R--Cu1--R group is of predominantly
ferromagnetic character. (We remind that our calculation has been done
in the room-temperature structure, whereas at 220~K the exchange interaction
does apparently change sign). Antiferromagnetic type of coupling over large
temperature range was observed in chemically related, but structurally
different, system Cu(hfac)$_2$L$^{\rm Pr}$, yielding $J$=$-$100~cm$^{-1}$
(see Ref.~\cite{ZhStCh43-163} for details). It is noteworthy that
all estimations of ``additional'' exchange parameters (those incorporating,
on the average, interactions to more distant Cu2 atoms and between
the chains) are by almost two orders of magnitude smaller.
This is consistent with relative unimportance of these more distant
interactions also according to the calculation. 

\begin{figure}
\centerline{\includegraphics[angle=0,width=12.5cm]{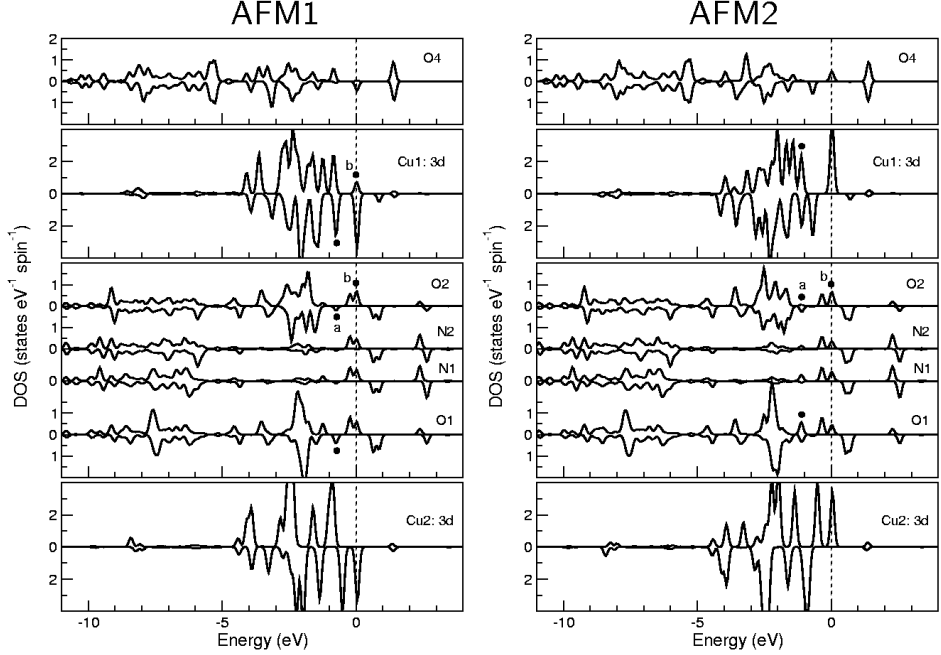}}
\caption{Spin-resolved partial DOS in two AFM configurations.
Zero energy separates occupied and unoccupied states.
The features marked by thick dots and labeled a and b are explained
in the text.
}
\label{fig:DOS}
\end{figure}

An inspection of partial DOS helps to understand
the origin of energy preference of the AFM2 configuration. 
Fig.~\ref{fig:DOS} compares the partial DOS at Cu1 and its neighbours
which exhibit a noticeable magnetization, and that of more distant Cu2.
For better comprehensibility, a very structured DOS from nondisperive
energy bands was broadened with half-width parameter of 0.1 eV. 
Obviously this completely smears the above-mentioned tiny band gap. 
Although formally dealing with a non-metal,
we refer to the energy which separates occupied and unoccupied states
as the Fermi energy (zero in Fig.~\ref{fig:DOS}).

Somehow simplifying, but in accord with chemical intuition, the Cu(II)
ion has its $3d$ majority-states fully occupied, and in the minority-spin
one state vacant. This vacant $3d$ state is responsible for a narrow peak 
in the DOS `at' (in fact, just above) the Fermi level.
This simple picture seems not quite consistent with nominal magnetic moments 
at Cu1 and Cu2 sites as shown in Table~\ref{tab:charge}, but that's merely because the Cu magnetization spills over its neighbours. It is well seen 
in spin density maps (not shown here) and in the maps of Cu-centered 
molecular orbitals shown below; moreover it is demonstrated by a presence 
of the Cu-related peak just above the Fermi level in the O4 DOS.
Apart from this, we have a magnetic moment of the radical, roughly coming from
summing up the values in Table~\ref{tab:charge} over O1, N1, N2 and O2
atoms where this moment is distributed (with a pronouncedly larger
portion residing on the more Cu-distant O1).

\begin{figure}
\centerline{\includegraphics[angle=0,width=12.8cm]{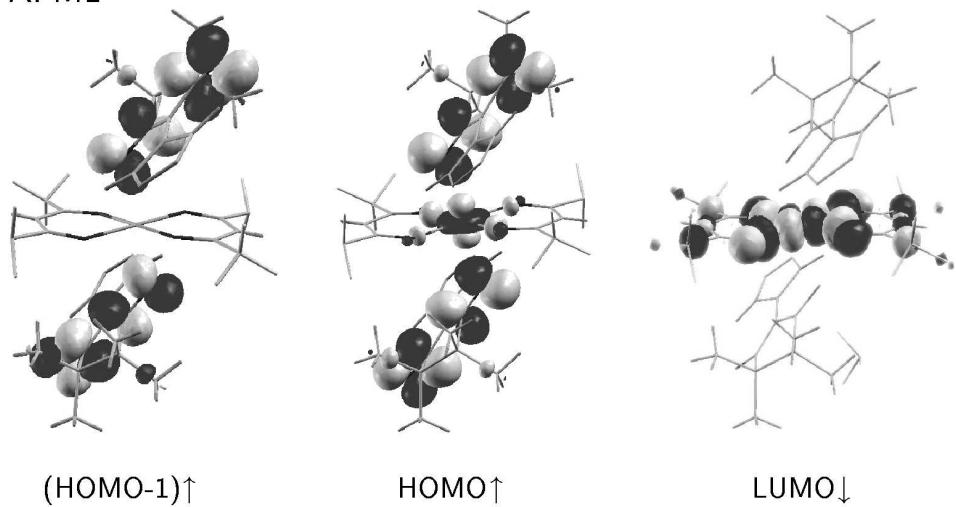}}
\caption{Highest occupied and lowest unoccupied molecular orbitals 
corresponding to ${\bm k}$=0 in the AFM1 configuration.
Only the Cu(hfac)$_2$ block and its two neighbouring ligands are shown.
Positive and negative values are indicated by two grayscale levels. 
As in an AFM configuration all orbitals are doubly degenerate, and the figures
show the vicinity of a selected Cu1 atom, the spin label is given for those 
respective orbitals which give a major contribution at the site in question,
consistently with Fig.~\ref{fig:DOS}. The opposite-spin
orbitals at the same energy are prominent at different Cu1 atom  
in the other polymer chain (see Fig.~\ref{fig:AFM}).
The plots are created with the XCrySDen program\cite{xcrysden}. 
}
\label{fig:AFM1-WF}
\end{figure}

\begin{figure}
\centerline{\includegraphics[angle=0,width=12.8cm]{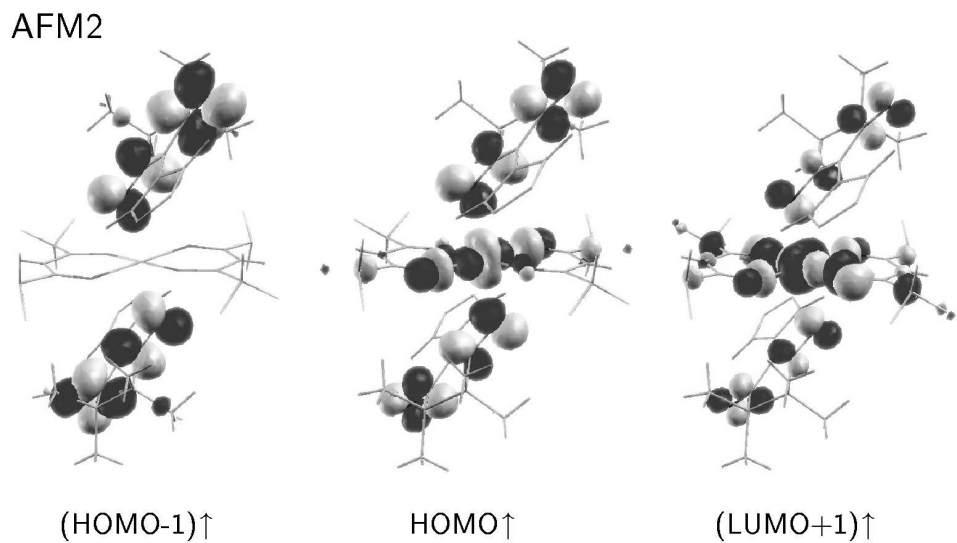}}
\caption{The same as in Fig.~\ref{fig:AFM1-WF}, for the AFM2 configuration.
}
\label{fig:AFM2-WF}
\end{figure}

As a most coarse observation, the Cu1-DOS is `inverted' when changing
from AFM1 to AFM2, but this is accompanied by a noticeable reconstruction
of electronic states, responsible to the Cu$3d$--O$2p$ hybridization.
Among Cu-related and R-related orbitals, which share the energy
interval of $-$4 to $+$1 eV around the Fermi level, we further concentrate
on some, with particular character of bonding.
The highest-energy peak among the fully occupied ones in majority-spin panel 
on the Cu1 site (spin-up for AFM1, spin-down for AFM2) comes from a state
which is \emph{nonbonding} with respect to the radical and antibonding
with respect to oxygens of the Cu(hfac)$_2$ group (note its presence
in the O4 spin-up DOS). The inversion of Cu1 spin changes merely the
position of this peak: from $-$0.83 eV in AFM1 to $-$0.68 eV in AFM2.
Further on, there is a state nonbonding with respect to Cu1 and pinned 
at the radical. It emerges in the DOS of O1, N1, N2, O2 as the lower 
of two peaks crossed by the Fermi level. Its wavefunction is shown in 
Figs.~\ref{fig:AFM1-WF} and \ref{fig:AFM2-WF} as that of the last-but-highest
occupied molecular orbital (-HOMO). It contains no contribution from Cu1 
whatsoever (having in fact a node at the Cu1 position) and is almost identical 
for AFM1 and AFM2 cases.
Then there are two pronounced states which involve both Cu1 and the radical:
(a) a feature in the minority-spin DOS at the Cu1 site, i.e., in the
spin-down channel of AFM1 (at $-$0.71 eV) and in the spin-up channel of AFM2
(at $-$1.13 eV); 
(b) the `just occupied' part of the highest of those two peaks
which make a distinguished two-peak structure best seen in the local DOS 
of the radical atoms, crossed by the Fermi level.
In fact this is the HOMO. This state is antibonding with respect to 
Cu1--O2 interaction. These (a) and (b) features are marked by thick dots
in the DOS figures.

The lowest unoccupied molecular orbital (LUMO) centered at the selected Cu1 
site corresponds to spin-down channel (the spin-up LUMO resides at a Cu1 atom 
in the other polymer chain) and contains no contribution from R. This is 
understandable, because within $\pm$0.5 eV around the Fermi level there is 
a gap in the spectrum of spin-down states of the radical.
The structure of the Cu$3d$ wavefunction within the LUMO is of the $xy$ type, 
in contrast to spin-up $z^2$ in the HOMO.
The next to LUMO (and higher by 16 meV) is the (also spin-down) 
orbital centered at Cu2.

When we now turn to the molecular orbitals of the AFM2 configuration 
(Fig.~\ref{fig:AFM2-WF}) we notice that the last-but-HOMO orbital, 
nonbonding to Cu1 and antisymmetric relative to its position,
has the same shape as for AFM1. However, the HOMO differs from that 
of the AFM1 case by its symmetry at the Cu1 site. It bounds two O4 atoms 
of the Cu(hfac))$_2$ group much stronger than two others; this contributes
to much higher disproportionality of magnetic moments over O4 atoms
in AFM2 case than in AFM1 (see Table \ref{tab:charge}). The LUMO
of AFM2 does not reflect the Cu1 states nor those of the radical,
but instead the Cu2 (see below). The next-to-LUMO contains again
the Cu1 contribution in somehow similar shape as in AFM1-LUMO$\downarrow$,
spread over the Cu(hfac)$_2$ plane. However -- and this is crucial --
the discussed wavefunction spreads over the O2--N2$\cdots$N1--O1 group,
in antibonding setting relative to Cu1.

As a result we see markedly stronger interaction of Cu$3d$ states
with the magnetic states of the radical in the AFM2 configuration than
in AFM1. This is so because the responsible states which tend to interact,
based on their spatial shape, belong now to the same spin channel,
namely to spin-up in the notation of Fig.~\ref{fig:DOS}.
As a single manifestation of this effect, the Cu-R hybridized state (a),
on the one side, and (b), but also the `(LUMO+1)$\uparrow$' of 
Fig.~\ref{fig:AFM2-WF}, on the other side, move apart. A similar argument 
applies with respect to other hybridized
Cu1--R states, resulting in a slight overall (albeit non-uniform)
replacement of occupied bands downwards. As a result, the AFM2
configuration wins over AFM1 due to the gain in band energy; moreover,
this accounts for a slightly larger band gap in AFM2.

The Cu2-DOS (the bottom panels in Fig.~\ref{fig:DOS} seems largely unaffected
by a reversal of its magnetic moment: its $3d_{\uparrow(\downarrow)}$ DOS 
in AFM2 is almost identical to $3d_{\downarrow(\uparrow)}$ one in AFM1.
There are tiny differences responsible for the coupling of Cu2 to the
`tail' of the ligand via N4 (see Fig.~\ref{fig:unit}), but they are left
beyond the scope of the present study.
It suffices to mention that the interaction with differently magnetized
Cu2 changes the sequence of orbitals in the vicinity of the Fermi level,
which are anyway nearly degenerate. In the AFM1 configuration,
the LUMO has a pronounced contribution at the Cu1 site (shown in the right
panel of Fig.~\ref{fig:AFM1-WF}), with the spin-down orientation, and the
next state (higher by 16 meV) is that centered at Cu2. In the AFM2 
configuration, the state with Cu2 predominance becomes the LUMO and have
no contribution at Cu1, whereas the Cu1-centered orbital is now the next
after LUMO (by 4 meV higher, shown in the right panel 
of Fig.~\ref{fig:AFM2-WF}), now evidently in the spin-up channel.

Summarizing, we investigated electronic structure, local densities of states
and the shape of molecular orbitals responsible for chemical bonding and
magnetic interaction in a heterospin polymer, connecting Cu(II) and
nitronyl nitroxide radical in a `head to head' configuration. We reproduced
in the calculation and explained at microscopic level the preference for
antiferromagnetic coupling between Cu and radical spins, and made an estimate
of the magnitude of this coupling. There are certain issues which seem
appealing for subsequent study. First, one should elucidate the interaction
between Cu2 ion coupled to the `tail' of the ligand and the presently
analyzed `exchange cluster'. Second, the variation of all interaction
parameters with temperature, according to (available by now)
crystallographic data\cite{Romanenko}. This might help to understand 
an unusual switch in the character of magnetic interaction, apparently 
accompanying a structural transition at 220 K described in 
Ref.~\cite{PhSS45-1465,RuChBull53-2406}.  

\section*{Acknowledgements}
The authors thank the Deutsche Forschungsgemeinschaft for financial support
(Priority Program SPP 1137 ''Molecular Magnetism''), and S.~V.~Fokin
for providing the crystallographic data of the compound studied.
AVP thanks E.~Z.~Kurmaev for introduction into the subject, 
and V.~I.~Ovcharenko and G.~V.~Romanenko for useful discussions.
AVG appreciates the funding by the Research Council of the President of the
Russian Federation (Grant NSH-1026.2003.2) and Russian Science
Foundation for Basic Research (Project 05-03-32707-a).

\end{document}